\documentclass[12pt]{article}
\usepackage{ amscd, amssymb, cite}
\usepackage[mathscr]{eucal}

\newcommand{\gen}[1]{{\left< #1 \right>}}
\newcommand{\ep}{\hfill $\blacksquare$}
\newcommand{\eep}{\hfill $\square$}
\newcommand{\pf}{\noindent {\bf Proof. \ }}

\newtheorem{thm}{Theorem}

\newtheorem{cor}{Corollary}
\newtheorem{defn}{Definition}
\newtheorem{ex}{Example}
\newtheorem{rem}{Remark}

\def\F{{\mathbb F}}
\def\I{{\mathcal I}}

\begin{document}

\baselineskip=18pt
\date{}
\title{\bf A note on complete classification of $(\delta+\alpha u^2)$-constacyclic codes of length $p^k$ over $\F_{p^m}+u\F_{p^m}+u^2\F_{p^m}$}
\author{Reza Sobhani$^{1,2}$, Zhonghua Sun$^3$, Liqi Wang$^3$, Shixin Zhu$^3$\\}
\maketitle\vspace{-.8cm}
\noindent {\scriptsize\center 1: Department of Mathematics, University of Isfahan, 81746-73441 Isfahan, Iran.\\
2: School of Mathematics, Institute for Research in Fundamental Sciences (IPM), 19395-5746 Tehran, Iran.\\
3: School of Mathematics, Hefei University of Technology, Hefei 230009, Anhui, P.R.China.\\}
\footnote{E-mail addresses: r.sobhani@sci.ui.ac.ir (R.~Sobhani), sunzhonghuas@163.com (Z.~Sun), liqiwangg@163.com (L.~Wang), zhushixin@hfut.edu.cn (S.~Zhu).}


\begin{abstract}

For units $\delta$ and $\alpha$ in $\F_{p^m}$, the structure of
$(\delta+\alpha u^2)$-constacyclic codes of length $p^k$ over
$\mathbb{F}_{p^m}+u\mathbb{F}_{p^m}+u^2\mathbb{F}_{p^m}$ is
studied  and self-dual $(\delta+\alpha u^2)$-constacyclic codes
are analyzed.
\end{abstract}


{\bf Keywords:} Constacyclic codes, Self-dual codes, Torsion codes.


\section{Introduction}

The class of constacyclic codes is an important class of linear
codes in coding theory, which can be viewed as a generalization of
cyclic codes. Let $R$ be a finite chain ring, a nonempty subset
$C$ of $R^n$ is called a linear code of length $n$ over
$R$ if $C$ is an $R$-submodule of $R^n$. For any unit
$\lambda$ in the ring $R$, $C$ is said to be
$\lambda$-constacyclic if $(\lambda c_{n-1},c_0,\dots,c_{n-2})\in
C$ for all $(c_0,c_1,\dots,c_{n-1})\in C$. If
an $n$-tuple $(c_0,c_1,\dots,c_{n-1})$ is identified with the
polynomial $c_0+c_1x+\dots+c_{n-1}x^{n-1}$ in the ring $R[x]/\gen{
x^n-\lambda }$, then $\lambda$-constacyclic codes of a given
length $n$ over $R$ are in correspondence with ideals of the ring
$R[x]/\gen{ x^n-\lambda }$. Special classes of repeated-root
constacyclic codes over certain finite chain rings have been
studied by numerous authors.

For any $s\geq 2$, let $R$ be the ring $\F_{p^m}[u]/\gen{u^s}$.
The ring $R$ has been widely used as alphabets in certain
constacyclic codes (See, for example, \cite{1,2,7,9,10,12,14} and
the references therein). In general, it seems to be difficult  to
classify all constacyclic codes over $R$ and only some constacyclic
codes of certain lengths are classified yet. Dinh \cite{4} classified all
constacyclic codes of length $2^s$ over the Galois extension rings
of $\F_2[u]/\gen{u^2}$ and gave their detailed structure. Later,
he classified and gave all constacyclic codes of length $p^s$ over
$\F_{p^m}[u]/\gen{u^2}$ in \cite{5}. Recently, Dinh et al.
\cite{8} classified negacyclic codes of $2p^s$ over
$\F_{p^m}[u]/\gen{u^2}$, analyzed the form of dual codes due to each type
and identified self-dual constacyclic codes. Chen
et al. \cite{3} classified all constacyclic codes of length $2p^s$
over $\F_{p^m}[u]/\gen{u^2}$. Also $(1+\alpha u)$-constacyclic codes of
arbitrary length $n$ over $\F_p[u]/\gen{u^s}$ have been studied in
details in \cite{9}, where $\alpha$ is a unit element in
$\F_p[u]/\gen{u^s}$.

From the above studies, we could see that
little work had been done on repeated-root $(\delta +\alpha
u^i)$-constacyclic codes over $R$, where $s\ge 3$ and $i\geq 2$.
Recently, the first author considered $(\delta +\alpha
u^2)$-constacyclic codes of length $p^k$ over
$\F_{p^m}[u]/\gen{u^3}$ and classified all $(\delta +\alpha
u^2)$-constacyclic codes of length $p^k$ over such ring in
\cite{13}. However, we have discovered a wrong conclusion claiming
that a certain polynomial $g(x)$ is constant, in the proof of
Theorem 8 in \cite{13}. Even though, that polynomial may not be
constant, the results before that theorem in \cite{13} are all correct
and those in the continuation of that theorem are correct only
for all constant such polynomials. Hence the classification of the
corresponding codes (for the non-constant case) and specially self-dual
codes, would be still incomplete. In this paper we
review some of the results presented
in \cite{13} and provide a complete classification for $(\delta
+\alpha u^2)$-constacyclic codes of length $p^k$ over
$\F_{p^m}[u]/\gen{u^3}$. We also try to classify self-dual such
codes and obtain some results in this respect.


\section{$(1+\alpha u^2)$-Constacyclic codes}

In this section we take a review to some structural results
presented in \cite{13}. Let $R$ be the ring
$\F_{p^m}[u]/\gen{u^3}$, $\alpha$ be a nonzero element in
$\F_{p^m}$ and $S$ be the ring $R[x]/\langle x^{p^k}-(1+\alpha
u^2)\rangle$. Let $\overline{S}$ be the ring $\F_{p^m}[x]/\langle
x^{p^k}-1 \rangle$ and $\mu:S\longrightarrow\overline{S}$ be the
map which sends $f(x)$ to $f(x)\bmod u$. For an ideal $C$ in $S$
and $0\le i\le 2$, define ${\rm Tor}_i(C)$ to be $\mu(\{f(x)\in S |
u^if(x)\in C \})$ which is an ideal of the ring $\overline{S}$ and
we call it the $i$-th torsion code of $C$. Clearly we have
${\rm Tor}_0(C)\subseteq {\rm Tor}_1(C)\subseteq {\rm Tor}_2(C)$.
Hence, related to $C$, there are integers $a\ge b\ge c$ such that
${\rm Tor}_0(C)=\gen{(x-1)^a}$, ${\rm Tor}_1(C)=\gen{(x-1)^b}$ and
${\rm Tor}_2(C)=\gen{(x-1)^c}$. The following theorem is a variation
of Theorem 2 in \cite{13}.
\begin{thm}\label{T1}\rm
Let $C$ be an ideal of $S$ and $a,b,c$ are as above. Then $C$ is
uniquely generated by the polynomials
\begin{eqnarray*}
f_0(x)=(x-1)^a+u(x-1)^tg(x)+u^2h(x),\\
f_1(x)=u(x-1)^b+u^2r(x),\\
f_2(x)=u^2(x-1)^c,
\end{eqnarray*}
in $C$, where $g(x), h(x)$ and $r(x)$ are polynomials in $\F_{p^m}[x]$ with
the property that $g(x)=0$ or $g(x)$ is a unit in $\overline{S}$
with $t+{\rm deg}(g(x))<b$, $r(x)=0$ or
${\rm deg}(r(x))<c$ and $h(x)=0$ or ${\rm deg}(h(x))<c$.\eep
\end{thm}
If $C$ is an ideal with the unique generators described in Theorem \ref{T1},
we write $C=\gen{\gen{f_0(x),f_1(x),f_2(x)}}$. Let $\I$ be the set of all ideals
of $S$. In \cite{13}, the set $\I$ has been divided to the following subsets.
\begin{eqnarray*}
{\mathcal A}:=\{C\in {\mathcal I}\ |\ c=0\ {\rm and}\
a+b\le p^k \},\\
{\mathcal A}':=\{C\in {\mathcal I}\ |\ c=0\ {\rm and}\
a+b\ge p^k \},\\
{\mathcal B}:=\{C\in {\mathcal I}\ |\ a=p^k\ {\rm and}\
b+c\le p^k \},\\
{\mathcal B}':=\{C\in {\mathcal I}\ |\ a=p^k\ {\rm and}\
b+c\ge p^k \},\\
{\mathcal C}:=\{C\in{\mathcal I} \ |\ a<p^k,\ c>0\ {\rm and}\
a+c\le p^k\},\\
{\mathcal C}':=\{C\in{\mathcal I} \ |\ a<p^k,\ c>0\ {\rm and}\
a+c\ge p^k\}.
\end{eqnarray*}
We refer to \cite{13} for the structure of ideals in ${\mathcal
A}\cup{\mathcal A}'\cup{\mathcal B}\cup{\mathcal B}'$. Here we only
review the structure of ideals in ${\mathcal C}\cup{\mathcal C}'$.
Let $C$ be an ideal of $S$. The annihilator of $C$,
denoted by $Ann(C)$, is defined to be the set
$\{f(x) | f(x)g(x)=0,\ for\ all\ g(x)\in C \}$. Clearly,
$Ann(C)$ is also an ideal of $S$. As declared in \cite{13},
the map $\eta:{\mathcal C}\longrightarrow{\mathcal C}'$
sending $C$ to $Ann(C)$ is a bijection and hence we only need
to determine ideals in ${\mathcal C}$. The following theorem is
indeed \cite[Theorem 7]{13}.
\begin{thm}\label{T2}\rm
Let $C=\gen{\gen{f_0(x),f_1(x),f_2(x)}}$ be an element of ${\mathcal C}$.
Then we must have the following:
\begin{itemize}
\item[(I)] $b=p^k-a+t$,
\item[(II)] $g(x)$ is a unit and $r(x)=\alpha g^{-1}(x)\bmod \gen{u,(x-1)^c}$,
\item[(III)] when $t\ne 2a-p^k-t$ we have $c\le {\rm min}\{t,2a-p^k-t\}$.\eep
\end{itemize}
\end{thm}
In Theorem 8 of \cite{13} it is claimed that for ideals in ${\mathcal C}$,
the polynomial $g(x)$ is a constant polynomial. This claim is not
true in general and $g(x)$ may be non-constant. In fact, the
conclusion made at lines 4 and 5 at page 131 in \cite{13} is wrong.
To be more accurate, in the following, we present a counter-example.
\begin{ex}\label{E1}\rm
Let $R=\F_{3}[u]/\gen{u^3}$, $S=R[x]/\gen{x^9-(1+u^2)}$ and
$$C=\gen{\gen{f_0(x),f_1(x),f_2(x)}},$$ where
\begin{eqnarray*}
f_0(x)=(x-1)^7+u(x-1)^2(1+(x-1))\\
f_1(x)=u(x-1)^4+u^2(1-(x-1))\\
f_2(x)=u^2(x-1)^2.
\end{eqnarray*}
It is easy to verify that $C$ is in the unique form and we are done.
\end{ex}
In the following theorem we provide a correct version of Theorems 8, 9 and 10 in \cite{13}.
\begin{thm}\label{T3}\rm
Let $C=\gen{\gen{f_0(x),f_1(x),f_2(x)}}$ be an element of ${\mathcal C}$. Then
\begin{itemize}
\item[1)] If $1\le c\le {\rm min}\{t,2a-p^k-t\}$ then we must have $\lceil(p^k+2)/2\rceil\le a\le p^k-1$, $1\le t\le 2a-p^k-1$, $a+c\le p^k$, $g(x)\in\overline{S}$ is a unit with ${\rm deg}(g(x))<p^k-a$, $r(x)=\alpha g^{-1}(x)\bmod (x-1)^c$ and $h(x)\in\overline{S}$ is a polynomial with ${\rm deg}(h(x))<c$.
\item[2)] If $c>{\rm min}\{t,2a-p^k-t\}$ then we must have $p=2$, $0\le t<2^{k-2}$, $a=t+2^{k-1}$, $t<c\le 2^{k-1}-t$, $g(x)\in\overline{S}$ is a unit with ${\rm deg}(g(x))<2^{k-1}-t$, $r(x)=\alpha g^{-1}(x)\bmod (x-1)^c$, $r(x)=g(x)\bmod (x-1)^{c-t}$ and $h(x)\in\overline{S}$ is a polynomial with ${\rm deg}(h(x))<c$.
\end{itemize}
Moreover, in both cases, if $g(x)r(x)=\alpha+(x-1)^cl(x)$ then we have $$Ann(C)=\gen{\gen{f'_0(x),f'_1(x),f'_2(x)}},$$ where $$f'_0(x)=(x-1)^{p^k-c}+u(x-1)^{a-c-t}[-r(x)]+u^2[-l(x)-(x-1)^{p^k-a-c}h(x)],$$ $$f'_1(x)=u(x-1)^{a-t}+u^2[-g(x)],$$ $$f'_2(x)=u^2(x-1)^{p^k-a}.$$
\end{thm}
\pf First, we assume that $c\le {\rm min}\{t,2a-p^k-t\}$. Note that, since $c\ge 1$, we can
not have $t=0$ and also we can not have $2a=p^k+t$. Hence $t\ge 1$ and $2a\ge
p^k+t+1\ge p^k+2$ and consequently $a\ge \lceil\frac{p^k+2}{2}\rceil$.
Also we have $1\le t\le 2a-p^k-1$. Other conditions comes from
Theorems \ref{T1} and \ref{T2}. Also, if $D:=\gen{f'_0(x),f'_1(x),f'_2(x)}$
one can easily conclude that $CD=0$ and hence similar arguments as those used
in \cite{13} shows that $D=Ann(C)$ and both $C$ and $D$ are in the unique form.
Now, if $c>{\rm min}\{t,2a-p^k-t\}$ then according to Theorem \ref{T2} we
must have $t=2a-p^k-t$ and hence $p=2$ and $a=2^{k-1}+t$.
Clearly, in this case, the representation of ${\rm Ann}(C)$ must has  the
form $\gen{\gen{F'_0(x),F'_1(x),F'_2(x)}}$, where
\begin{eqnarray*}
F'_0(x):=(x-1)^{2^k-c}+u(x-1)^{2^{k-1}-c}g'(x)+u^2h'(x),\\
F'_1(x):=u(x-1)^{2^{k-1}}+u^2r'(x),\\
F'_2(x):=u^2(x-1)^{2^k-a}.
\end{eqnarray*}
From $0=F_0(x)F'_1(x)$ we can conclude that $g'(x)=r(x)$ and from $0=F'_0(x)F_1(x)$
we can deduce that $r'(x)=g(x)$. Now, $0=F_0(x)F'_0(x)$ implies that
$r(x)=g(x)\bmod (x-1)^{c-t}$. Now, setting $D:=\gen{f'_0(x),f'_1(x),f'_2(x)}$,
the remaining of the proof is similar to that for the first part.
\ep
\begin{rem}\label{R1}\rm
As it was proved in [13], the above obtained results can be extended to $(\delta+\alpha u^2)$-constacyclic codes of length $p^k$ over $R$.
\end{rem}

\section{Self-dual $(1+\alpha u^2)$-constacyclic codes}

In this section we try to find Euclidean self-dual $(1+\alpha
u^2)$-constacyclic codes of length $p^k$. For two polynomials
$f(x)=\sum_{i=0}^{p^k-1}f_ix^i$ and
$g(x)=\sum_{i=0}^{p^k-1}g_ix^i$ in $S$ we define the inner
product of $f(x)$ and $g(x)$ to be
$$f(x)\cdot g(x)=\sum_{i=0}^{p^k-1}f_ig_i.$$ With respect to this
inner product, the dual of an ideal $C$ of $S$, denoted by
$C^{\perp}$, is defined as $$C^{\perp}=\{f(x)\in S\ |\ f(x)\cdot
g(x)=0\ {\rm for\ all\ } g(x)\in C\}.$$ The ideal $C$ is said to
be self-dual, if $C=C^{\perp}$. To determine self-dual codes we
need the following definition.
\begin{defn}\rm
Let $f(x)=\sum_{i=0}^{p^k-1}{a_ix^i}$ be an element of $S$. The
reciprocal of $f(x)$, denoted $f^*(x)$, is the element
$\sum_{i=0}^{p^k-1}{a_ix^{p^k-i}}$ of $S$. For any subset $E$ of
$S$, the set $\{e^*\ |\ e\in E\}$ is denoted by $E^*$.
\end{defn}
We know from \cite[Lemma 1]{13} that $C^{\perp}={\rm Ann}(C)^*$.
Also, we know from \cite{13} that, self-dual $(1+\alpha
u^2)$-constacyclic codes exist only in the case $p=2$ and the only
self-dual ideal in ${\mathcal A}\cup{\mathcal A}'\cup{\mathcal
B}\cup{\mathcal B}'$ is the ideal $\gen{\gen{u(x-1)^{2^{k-1}}, u^2
}}$.

Now, let $C$ be a self-dual ideal in ${\mathcal C}$. Since we must
have $$2^k-a+t=T_1(C)=T_1(C^{\perp})=T_1({\rm Ann}(C))=a-t,$$
hence we have $t=a-2^{k-1}$. Also we have
$$a+c=T_0(C)+T_2(C)=T_0(C^{\perp})+T_2(C^{\perp})=T_2({\rm Ann}(C))+T_0({\rm Ann}(C))=2^k-a+2^k-c,$$
and hence $a+c=2^k$. Therefore a self-dual ideal $C$ must belong
to ${\mathcal C}\cap{\mathcal C'}$. By now, a self-dual ideal $C$
in $S$ must has the representation
$C=\gen{\gen{f_0(x),f_1(x),f_2(x)}}$, where
$$f_0(x)=(x+1)^{2^{k-1}+t}+u(x+1)^tg(x)+u^2h(x),$$
$$f_1(x)=u(x+1)^{2^{k-1}}+u^2r(x),$$ $$f_2(x)=u^2(x+1)^{2^{k-1}-t},$$  $0\le
t<2^{k-1}$, $g(x)\in \overline{S}$ is invertible with ${\rm
deg}(g(x))<2^{k-1}-t$, $r(x)=\alpha g^{-1}(x)\bmod
(x+1)^{2^{k-1}-t}$ and $h(x)\in\overline{S}$ is a polynomial with
${\rm deg}(h(x))<2^{k-1}-t$. Moreover if $t<2^{k-2}$ then,
additionally, we must have $g(x)=r(x)\bmod (x+1)^{2^{k-1}-t}$.
From now on, we denote such an ideal $C$ with
$Ideal(k,t,g(x),h(x))$. It can be deduced from $C^{\perp}={\rm
Ann}(C)^*$ and some simple calculations that
$C^{\perp}=Ideal(k,t,G(x),H(x))$, where
$$G(x)=(\alpha g^{-1}(x))^*\bmod (x+1)^{2^{k-1}-t},$$
$$H(x)=[\alpha(x+1)^t+x^{2^{k-1}+t}h^*(x)+q(x)g^*(x)]\bmod
(x+1)^{2^{k-1}-t},$$ and $q(x)$ is such that $(\alpha
g^{-1}(x))^*=(x+1)^{2^{k-1}-t}q(x)+G(x)$. As a consequence of the
above discussions and with the same notation, we have the
following theorem.
\begin{thm}\label{T4}\rm
Let $k$ be a positive integer, $0\le t<2^{k-1}$ and
$C=Ideal(k,t,g(x),h(x))$ be an element of $S$. Then $C$ is a
self-dual ideal, if and only if, all the following conditions
hold.
\begin{itemize}
\item[1)] $g(x)g^*(x)=\alpha\bmod(x+1)^{2^{k-1}-t}$.

\item[2)] if $t<2^{k-2}$, then
$g^2(x)=\alpha\bmod(x+1)^{2^{k-1}-2t}$.

\item[3)] $H(x)=h(x)$.\eep
\end{itemize}
\end{thm}
The first step to find self-dual ideals of $S$ is to find
polynomials $g(x)\in\F_{p^m}[x]$ for which we have ${\rm
deg}(g(x))<2^{k-1}-t$ and
$g(x)g^*(x)=\alpha\bmod(x+1)^{2^{k-1}-t}$. Clearly, the constant
polynomial $g(x)=\sqrt{\alpha}$, considered in \cite{13}, is one
of such polynomials. Write $g(x)=\sum_{i=0}^{2^{k-1}-t-1}g_i(x+1)^i$.
Then we have
\begin{eqnarray*}
g^*(x) & = & \sum_{i=0}^{2^{k-1}-t-1}g_i(x+1)^ix^{2^{k}-i}\\
       & = & \sum_{i=0}^{2^{k-1}-t-1}g_i(x+1)^i\sum_{j=0}^{2^k-i}{{2^k-i}\choose j}(x+1)^j\\
       & = & \sum_{i=0}^{2^{k-1}-t-1}\left(\sum_{j=0}^{i}{{2^k-j}\choose i-j}g_j\right)(x+1)^i\bmod (x+1)^{2^{k-1}-t}.
\end{eqnarray*}

For $0\le s\le 2^{k-1}-t-1$, set $\sigma_s:=\sum_{i=0}^{s}\sum_{j=0}^{s-i}{{2^k-j}\choose s-i-j}g_ig_j$. The first condition in Theorem \ref{T4} now becomes
$\sigma_0=\alpha$, i.e. $g_0=\sqrt{\alpha}$, and for $1\le s\le 2^{k-1}-t-1$, $\sigma_s=0$. This is a nonlinear system of equations and solving it is not an easy task! To estimate the form of solutions, let us assume that $g(x)$ has the form $g_0+g_1(x+1)+g_2(x+1)^2$, i.e. $g_i=0$ for $i\ge 3$. In this case, the equations become
\begin{eqnarray*}
\left\{
  \begin{array}{ll}
    \sigma_0=g^2_0=\alpha \\
    \sigma_1=0\\
    \sigma_2=\sigma_3=g_1(g_0+g_1)=0\\
    \sigma_4=(g_0+g_1)(g_1+g_2)+g_2^2=0\\
    \ \ \ \ \ \ \ \vdots
  \end{array}
\right.
\end{eqnarray*}
Therefore, if $2^{k-1}-t\ge 5$ then $g(x)=\sqrt{\alpha}$, $g(x)=\sqrt{\alpha}x$ and $g(x)=\sqrt{\alpha}x^2$ are the only solutions for the system of equations.
Other solutions corresponding to the case $2^{k-1}-t\le 4$, rather than those of the form given above, are as follows:
\begin{eqnarray*}
\begin{array}{lll}
2^{k-1}-t=2 & \Longrightarrow & g(x)=\sqrt{\alpha}+\beta (x+1)\\
2^{k-1}-t\in\{3,4\} & \Longrightarrow & g(x)=\sqrt{\alpha}+\beta (x+1)^2\ {\rm or}\ g(x)=\sqrt{\alpha}x+\beta (x+1)^2,
\end{array}
\end{eqnarray*}
where, $\beta\in\F_{2^m}$. As we mentioned, obtaining all solutions are difficult but we can now conjecture that, if we restrict the degree of $g(x)$ to the positive integer $d$ then there exist a positive integer $D$ depending on $d$ such that for all $k,t$ for which $2^{k-1}-t>D$, the polynomials $g(x)=\sqrt{\alpha}x^s$, are the only solutions for the system of equations. From this motivation, in what follows, we classify self-dual codes for which the polynomial $g(x)$ has the form $g(x)=\sqrt{\alpha}x^s$, for some $0\le s \le 2^{k-1}-t$. Let $w=w(s)$ be such that when $s\ne 0$, $2^w\|s$ and when $s=0$, $w=k-2$. We now have the following theorem.
\begin{thm}\label{T5}\rm
Let $k$ be a positive integer, $0\le t<2^{k-1}$, $0\le
s<2^{k-1}-t$, $h(x)\in\overline{S}$ with ${\rm deg}(h(x))<2^{k-1}-t$ and $C=Ideal(k,t,\sqrt{\alpha}x^s,h(x))$ be an
element of $S$. Then $C$ is a self-dual ideal, if and only if, one
of the following conditions hold.
\begin{itemize}
\item[1)] $2^{k-2}\le t<2^{k-1}$ and
$x^{2^{k-1}+t}h^*(x)=h(x)\bmod(x+1)^{2^{k-1}-t}$.

\item[2)] $2^{k-2}-2^w\le t<2^{k-2}$ and
$[\alpha(x+1)^t+x^{2^{k-1}+t}h^*(x)]=h(x)\bmod(x+1)^{2^{k-1}-t}$.
\end{itemize}
\end{thm}
\pf The first part follows from Theorem \ref{T4} and the facts
that $\alpha(x+1)^t=0\bmod (x+1)^{2^{k-1}-t}$ and $q(x)=0$. For
the second part, note that if $t<2^{k-2}$ then according to the
second part of Theorem \ref{T4} we must have $g^2(x)=\alpha\bmod
(x+1)^{2^{k-1}-2t}$ and hence we must have $(x+1)^{2^{k-1}-2t}\mid
(x^{2s}+1)$. But, if we write $s=2^ws'$ with $s'$ odd, then we
have $x^{2s}+1=(x^{s'}+1)^{2^{w+1}}$. Hence
$(x+1)^{2^{k-1}-2t}\mid (x^{2s}+1)$ if and only if $2^{k-1}-2t\le
2^{w+1}$ or equivalently, $t\ge 2^{k-2}-2^w$. The remaining of the
proof follows from the third part of Theorem \ref{T4} and the fact
that $q(x)=0$. \ep
\begin{cor}\label{C1}\rm
Let $k$ be a positive integer, $0\le t<2^{k-1}$ and $1\le
s<2^{k-1}-t$. If $C=Ideal(k,t,\sqrt{\alpha}x^s,h(x))$ is a
self-dual ideal then we must have $s\le 2^{k-2}$.
\end{cor}
\pf If $t\ge 2^{k-2}$ then clearly we must have $s<2^{k-2}$.
If $t<2^{k-1}$ then according to the second part of Theorem
\ref{T5}, we must have $t\ge 2^{k-2}-2^w$, where $2^w\| s$.
But if $s>2^{k-2}$ then we have $2^{k-2}-2^w>2^{k-1}-s-1$
implying $s+t>2^{k-1}-1$ which is a contradiction. Hence we
always must have $s\le 2^{k-2}$ and the proof is completed.\ep

Let us denote by $N(2^k,t,s)$ the number of self-dual ideals
of $S$ of the form $C=Ideal(k,t,\sqrt{\alpha}x^s,h(x))$, where
$$h(x)=h_0+h_1(x+1)+\cdots+h_{2^{k-1}-t-1}(x+1)^{2^{k-1}-t-1}$$
is a polynomial in $\overline{S}$. Recall that $M(2^k,t)$ is the following
matrix defined in \cite{11,13}:
\begin{eqnarray*}
M(2^k,t):= \left(
\begin{array}{cccccc}
  0 & 0 & \dots &\dots& 0 \\
  {{2^{k-1}+t}\choose 1} & 0 & \ddots & \ddots & 0 \\
  {{2^{k-1}+t}\choose 2} & {{2^{k-1}+t-1}\choose 1} &  \ddots & \ddots & 0 \\
  \vdots & \vdots & \mbox{} & \ddots & \vdots \\
  {2^{k-1}+t}\choose {2^{k-1}-t-1} & {2^{k-1}+t-1}\choose {2^{k-1}-t-2} & \dots &\dots &  0 \\
\end{array}
\right).
\end{eqnarray*}
Now, from Theorem \ref{T5}, we have $C=Ideal(k,t,\sqrt{\alpha}x^s,h(x))$ is self-dual, if and only if
\begin{equation}\label{eq1}
u^2M(2^k,t)(h_0,h_1,\cdots,h_{2^{k-1}-t-1})^{tr}=u^2(d_0,d_1,\cdots,d_{2^{k-1}-t-1})^{tr},
\end{equation}
where $d_i=\alpha$ if $i=t$ and $d_i=0$ otherwise. But
Equation (\ref{eq1}) has solutions for
$(h_0,h_1,\cdots,h_{2^{k-1}-t-1})$, if and only if
\begin{equation}\label{eq2}
M(2^k,t)(h_0,h_1,\cdots,h_{2^{k-1}-t-1})^{tr}=(d_0,d_1,\cdots,d_{2^{k-1}-t-1})^{tr},
\end{equation}
has solutions for $(h_0,h_1,\cdots,h_{2^{k-1}-t-1})$ in $\F_{2^m}$. Now we
have the following theorem:
\begin{thm}\label{T11}\rm
Let ${\mathcal K}$ be the nullity of $M(2^k,2^{k-1}-t)$ over $\F_{2^m}$
Then we have
\[N(2^k,t,s)=\left\{%
\begin{array}{ll}
    (2^m)^{\mathcal K},   & \mbox{ when there is a solution for $(h_0,h_1,\dots,h_{2^{k-1}-t-1})^{tr}$;} \\
    0,         & \mbox{ otherwise.}
\end{array}%
\right.\]\eep
\end{thm}
On the other hand, it has been proved in \cite{15} that Equation
(\ref{eq2}) always has solutions for $(h_0,h_1,\cdots,h_{2^{k-1}-t-1})$ in
$\F_{2^m}$ except when $t=0$. Also it has been shown there
that ${\mathcal K}=\lceil\frac{2^{k-1}-t+1}{2}\rceil$. Therefore we have
\[N(2^k,t,s)=\left\{%
\begin{array}{ll}
    (2^m)^{\lceil\frac{2^{k-1}-t+1}{2}\rceil},   & \mbox{ when $t\ne 0$;} \\
    0,         & \mbox{ when $t=0$.}
\end{array}%
\right.\] Hence we have the following corollary.
\begin{cor}\rm
Let $N(2^k)$ denote the number of self-dual ideals of $S$ having the form $Ideal(k,t,\sqrt{\alpha}x^s,h(x))$. We have {\small
\begin{eqnarray*}
  N(2^k)=\sum_{s=0}^{2^{k-2}}\sum_{t={\rm max}\{2^{k-2}-2^w,1\}}^{2^{k-1}-s-1}(2^m)^{\lceil\frac{2^{k-1}-t+1}{2}\rceil}.
\end{eqnarray*}
}\eep
\end{cor}
As described before, when we restrict the degree of $g(x)$ to a non-negative integer $d$ then, for some limited values of $t$, there might exist some choises for $g(x)$ rather than those of the form $\sqrt{\alpha}x^s$. Let us complete the classification of self-dual ideals of $S$ for which $g(x)=g_0+g_1(x+1)+g_2(x+1)^2$ has degree at most $2$. We now only need to consider the following three cases
\begin{eqnarray*}
\begin{array}{llllll}
(I)  & 2^{k-1}-t=2 & {\rm and} & g(x)=\sqrt{\alpha}+\beta (x+1) & {\rm with}\ \beta\notin\{0,\sqrt{\alpha}\}.\\
(II) & 2^{k-1}-t=3 & {\rm and}, & g(x)=\sqrt{\alpha}+\beta (x+1)^2  & {\rm with}\ \beta\notin\{0,\sqrt{\alpha}\}\\
     &                     &            & {\rm or}                          &                                  \\
     &                     &            & g(x)=\sqrt{\alpha}x+\beta (x+1)^2 & {\rm with}\ \beta\ne 0.\\
(III) & 2^{k-1}-t=4 & {\rm and}, & g(x)=\sqrt{\alpha}+\beta (x+1)^2  & {\rm with}\ \beta\notin\{0,\sqrt{\alpha}\}\\
     &                     &            & {\rm or}                          &                                  \\
     &                     &            & g(x)=\sqrt{\alpha}x+\beta (x+1)^2 & {\rm with}\ \beta\ne 0.
\end{array}
\end{eqnarray*}
But, conditions 2 and 3 in Theorem \ref{T4} imply that there is no self-dual ideal in the cases $(I)$ and $(III)$ . Also, in the case $(II)$, an easy computation gives us the following self-dual ideals:\\
$$\gen{\gen{(x+1)^5+u(x+1)g(x)+u^2h(x),u(x+1)^4+u^2r(x),u^2(x+1)^3}},$$ where $g(x)=\sqrt{\alpha}+\beta(x+1)^2$, $h(x)=h_0+h_1(x+1)+h_2(x+1)^2$, $h_0=\alpha+\sqrt{\alpha}\beta+\beta^2$, and $\beta,h_1,h_2\in \F_{2^m}$ with $\beta\notin\{0,\sqrt{\alpha}\}$.\\
$$\gen{\gen{(x+1)^5+u(x+1)g(x)+u^2h(x),u(x+1)^4+u^2r(x),u^2(x+1)^3}},$$ where $g(x)=\sqrt{\alpha}x+\beta(x+1)^2$, $r(x)=g(x)+\sqrt{\alpha}(x+1)^2$, $h(x)=h_0+h_1(x+1)+h_2(x+1)^2$, $h_0=\beta^2+\alpha$ and $\beta,h_1,h_2\in \F_{2^m}$ with $\beta\ne 0$.\\
$$\gen{\gen{(x+1)^{2^k-3}+u(x+1)^{2^{k-1}-3}g(x)+u^2h(x),u(x+1)^{2^{k-1}}+u^2r(x),
u(x+1)^3}},$$ where $g(x)=\sqrt{\alpha}+\beta(x+1)^2$, $h(x)=h_0+h_1(x+1)+h_2(x+1)^2$, $h_0=\sqrt{\alpha}\beta+\beta^2$, $k\ge 4$ and $\beta,h_1,h_2\in \mathbb{F}_{2^m}$ with $\beta\notin\{0,\sqrt{\alpha}\}$.\\
$$\gen{\gen{(x+1)^{2^k-3}+u(x+1)^{2^{k-1}-3}g(x)+u^2h(x),u(x+1)^{2^{k-1}}+u^2r(x),u(x+1)^3}},$$
where $g(x)=\sqrt{\alpha}x+\beta(x+1)^2$, $h(x)=h_0+h_1(x+1)+h_2(x+1)^2$, $h_0=\beta^2$, $k\ge 4$ and $\beta,h_1,h_2 \in \F_{2^m}$  with $\beta\ne 0$.\\


\section{Acknowledgments}
The research of the first author was in part supported by a grant from IPM (No. 94050080).
The research of the third author was in part supported by Fundamental Research Funds through the Central Universities (No.JZ2015HGBZ0499, JZ2016HGXJ0089), and 
the research of the fourth author was in part supported by the National Natural Science Foundation of China(No.61370089). 

\end{document}